\documentclass[reprint,twocolumn,superscriptaddress,showpacs,nofootinbib,notitlepage,preprintnumbers]{revtex4-2}
\usepackage{placeins}
\usepackage{graphicx}
\usepackage{hyperref}
\usepackage[dvipsnames]{xcolor}
\usepackage{amsmath,amsthm,amssymb}
\usepackage{comment}

\usepackage{multirow}
\usepackage[normalem]{ulem}

\newcommand{\obs}{\boldsymbol{D}}
\newcommand{\params}{\boldsymbol{\alpha}}
\newcommand{\altvar}{\boldsymbol{\omega}}
\newcommand{\paramsOneD}{\alpha}
\newcommand{\altvarOneD}{\omega}

\begin{document}
\preprint{INT-PUB-23-041}

\title{Normalizing Flows for Bayesian Posteriors: Reproducibility and Deployment}

\author{Yukari Yamauchi}
\email{yyama122@uw.edu}
\affiliation{Institute for Nuclear Theory, University of Washington, Seattle, WA 98195, USA}

\author{Landon Buskirk}
\affiliation{Facility for Rare Isotope Beams, Michigan State University, East Lansing, Michigan 48824, USA}
\author{Pablo Giuliani}

\affiliation{Facility for Rare Isotope Beams, Michigan State University, East Lansing, Michigan 48824, USA}
\affiliation{Department of Statistics and Probability, Michigan State University, East Lansing, Michigan 48824, USA}
\author{Kyle Godbey}

\affiliation{Facility for Rare Isotope Beams, Michigan State University, East Lansing, Michigan 48824, USA}

\date{\today}

\begin{abstract}
We present a computational framework for efficient learning, sampling, and distribution of general Bayesian posterior distributions.
The framework leverages a machine learning approach for the construction of normalizing flows for the general probability distributions typically encountered in Bayesian uncertainty quantification studies.
This normalizing flow can map a trivial distribution to a more complicated one and can be stored more efficiently than the empirical distribution samples themselves.
Once the normalized flow is trained, it further enables parallelized and uncorrelated sampling of the learned distribution.
We demonstrate our framework with three test distributions with strong non-linear correlations, multi-modality, and heavy tails, as well as with a realistic posterior distribution obtained from a Bayesian calibration of a nuclear relativistic mean-field model.
The performance of the framework, as well as its relatively simple implementation, positions it as one fundamental cornerstone in the development and deployment of continuous calibration pipelines of physical models and as a key component of future reproducible science workflows.
\end{abstract}

\maketitle

\section{Introduction}\label{sec:introduction}
Uncertainty quantification (UQ) is a crucial aspect of scientific inquiry across various domains, such as physics, biology, engineering, and social sciences.
A driving motivation of UQ studies is to assess the credibility of experimental methods and measurements, numerical simulations, and physical models, and to rigorously model and propagate the uncertainties coming from all such sources.
UQ efforts often directly inform decision making based on the estimated uncertainties, such as optimizing for the best design of an apparatus, selecting the most suitable parameter region of a theoretical model, or evaluating the potential impact of an experimental campaign.

Given such an important and widespread application spectrum, a careful consideration of reproducibility and accessibility should reside at the heart of any UQ study.
Reproducibility means that the same results can be obtained by different researchers using the same data and methods, while accessibility means that the data and methods are available and understandable to others.
In practice, achieving reproducibility and accessibility in UQ studies can be challenging, especially when dealing with high-dimensional data sets and Bayesian posterior distributions, commonly featured in many scientific applications.
One possible way to compress and share the relevant probability distributions is to approximate them by multivariate Gaussian laws (see for example covariance analysis for nuclear Density Functional Theory models \cite{chen2014building,klupfel2009variations,erler2013energy}). This reduction, or simplification, comes at the cost of possible misrepresentation of the underlying distributions, ignoring features such as multi-modality, heavy tails, and non-linear relations between variables. Variational Bayes methods are another approach for representing posterior distributions (see \cite{jordan99,titsias2014doubly,blei2017variational,kejzlar2023black,kejzlar2023variational} for reviews and examples of applications), expanding the Gaussian law approximation by including a family of distributions with parameters to be determined by the data. Normalizing flows -realized through a machine learning approach- is another possibility, and the method we explore in this paper.

Normalizing flows can model arbitrary probability distributions under some reasonable assumptions~\cite{9089305} by transforming a simple base distribution, such as a standard multivariate normal distribution, into a more complex target distribution, using a sequence of invertible and differentiable functions.
By learning this mapping from the base distribution to the target distribution, normalizing flows can capture the complex dependencies and variations in the data, and provide a compact and expressive representation of the probability distribution without relying on a specific functional form. Modern machine learning methods provide a perfect computational workbench to effectively learn such flows from the available data (see, e.g.,~\cite{gabrié2021efficient,Rizvi:2023mws} for other applications in Bayesian analysis). 

In this work, we exploit the flexibility of normalizing flows to devise a framework of efficient compression and distribution of high-dimensional probability distributions.
Such a framework allows us to store and generate samples from arbitrary probability distributions, all at a minimal cost in terms of computing resources and development effort. We organize the rest of the manuscript as follows. In Section~\ref{sec: Formalism} we give an overview of Bayesian statistics, explaining how posterior distributions are constructed from physical models and data. In this section we also review the theory behind normalizing flows and outline the machine learning framework we use for training such maps. In Section~\ref{sec:demonstration} we demonstrate the effectiveness of the approach on analytic distributions as well as published datasets from nuclear theory.
We close in Section~\ref{sec:conclusion} with our conclusions and outlooks.

\section{Formalism}\label{sec: Formalism}

\subsection{Bayesian Statistics}

Bayesian statistics offers a probabilistic framework for making inferences based on observed data within the context of uncertainty. One of its main advantages is the explicit and clear statement of all assumptions concerning the various sources of uncertainties and how they interact. This explicitness allows for the transparent scrutiny and challenging of these assumptions, facilitating improvements in the developed statistical framework. Another major advantage is its capacity to continuously adapt knowledge based on new data, all within the language of probabilities. In the Bayesian framework, prior beliefs about model parameters and predictions are combined with observed data to produce a posterior distribution, effectively refining our understanding as new information becomes available. This process is constructed through Bayes' theorem \cite{gelman1995bayesian,phillips2021get}:
\begin{equation}\label{eq:post}
     P(\params|\obs) = \frac{P(\obs|\params) P (\params)}{P(\obs)}. 
\end{equation}
The \emph{posterior} distribution $P(\params|\obs)$ encodes the new information on the model parameters $\params$ -or any other quantities of interest- given the new observed data $\obs$. This posterior is constructed from the combination of the \emph{likelihood} $P(\obs|\params)$, which models how the observed data $\obs$ could have arisen for a given parameter configuration $\params$, and the prior $P (\params)$, which accounts for all the previous information and expert knowledge on the parameters $\params$. The \emph{evidence} $P(\obs)$ gives a sense on how well the current model, including the statistical treatment for the uncertainties, explains overall the data $\obs$. In many applications, this evidence term is often regarded as a normalization constant that makes the posterior a true normalized probability distribution.

The specific form of the posterior distribution~\eqref{eq:post} varies wildly across the many domains and applications of Bayesian statistics, adapting for each problem's specific characteristics. For relatively simple scenarios these posteriors could take explicit forms -such as multivariate Gaussians- that are easy to sample from and to calculate moments such as means and covariances. For more complicated scenarios where non-linear models or unknown error scales are involved, for example, obtaining information from the posterior can become a challenge. Markov-Chain Monte Carlo (MCMC) sampling techniques, such as the Metropolis Hastings algorithm and variations such as Hamiltonian Monte Carlo methods \cite{gelman1995bayesian}, are usually employed in these situations. Many of these sampling methods could require several chains (walkers) to go through thousands to millions of samples to fully explore the parameter space, with the situation turning more problematic the more parameters are involved (the curse of dimensionality), or if the distributions show multi-modality and high-correlation structures. 
The correlation function:
\begin{eqnarray}\label{eq:ac}
    C_i(\tau) &=& \Big\langle (\alpha_i(0) - \langle\alpha_i(0)\rangle)(\alpha_i(\tau) - \langle\alpha_i(\tau)\rangle) \Big\rangle \nonumber\\
    &\approx& \frac{1}{N}\sum_{t=1}^{N} (\alpha_i(t)-\langle\alpha_i\rangle)(\alpha_i(t+\tau)-\langle\alpha_i\rangle) \nonumber\\
    &\propto& e^{-\tau/l}
\end{eqnarray}
can be used to estimate the auto-correlation length $l$ of a parameter $\alpha_i$ in the chain. However a precise measurement of $l$ typically requires a large sampling size.

The usual requirement of such a vast number of model evaluations for estimating any quantity of interest is one of the main computational bottlenecks for proper uncertainty quantification studies, and one of the main motivations for the development and use of several model emulation strategies \cite{mcdonnell2015uncertainty,giuliani2023bayes,surer2023}. On the other hand, even when the sampling rate can be made fast enough, relying on storing the samples for later distribution and for future refining when new data becomes available quickly can become an inefficient process, specially when dealing with large sets of calculated (model) data. Normalizing flows, as we discuss in the following section, can serve to improve this process.

\subsection{Normalizing Flow}

A normalizing flow (NF) is a map $\mathbb{R}^N \rightarrow \mathbb{R}^N$ which induces a non-trivial distribution from a trivial distribution. In our context, NFs induce a desired
$N-$dimensional density $p(\params)$ from the $N$-dimensional Gaussian distribution $g_N(\altvar)$. For the rest of this work we will use the shorthand notation $p(\params)$ to represent any general  distribution, not necessarily normalized to integrate 1, a specific case being the un-normalized posterior $\propto P(\obs|\params)P (\params)$ from Eq.~\eqref{eq:post}.

In mathematical terms, the NF $f$ with $\params = f(\altvar)$ for the density $p(\params)$ satisfies the equation:
\begin{equation}\label{eq:nf_exact}
    d\params \; p(\params) = d\altvar\;\det\left( \frac{\partial \params}{\partial \altvar}  \right) p(f(\altvar)) = d\altvar\; g_N(\altvar),
\end{equation}
where $\det\left( \frac{\partial \params}{\partial \altvar}  \right)$ is the Jacobian matrix. Under reasonable assumptions, such NF exist for any normalizable (real and non-negative) distributions (see, e.g.,~\cite{9089305} and references therein). As illustration, in 1-dimension, a NF $\altvarOneD\rightarrow\paramsOneD$ for a distribution $\pi(\paramsOneD)$ of a single variable:
\begin{equation}
    \frac{d\paramsOneD}{d\altvarOneD} \pi(\paramsOneD) = g_1(\altvarOneD)
\end{equation}
can be constructed via the cumulative distribution functions:
\begin{equation}
    \Pi(\paramsOneD) = \int_{-\infty}^\paramsOneD d\paramsOneD'\;\pi(\paramsOneD') \;\;\mathrm{and}\;\;G_1(\altvarOneD) = \int_{-\infty}^\altvarOneD d\altvarOneD'\;g_1(\altvarOneD')\;\text.
\end{equation}

The cumulative distribution functions can be seen as a map from a probability distribution to the uniform distribution on a unit interval. Therefore a NF for $\pi$ is constructed as $f = \Pi^{-1} \circ G_1$, where ``$\circ$'' denotes composition . In higher dimensions, NFs are known to exist albeit they cease to be unique. Finding NFs with desirable properties is an active area of research (see~\cite{villani2003topics} for a review).

An exact NF, once constructed, enables us to obtain samples $\params$ from the original probability distribution $p(\params)$ by simply sampling $\altvar$ from the $N$-dimensional Gaussian distribution and applying the NF, i.e. $\params=f(\altvar)$. In the same way, the expectation value of any function $\mathcal{O}(\params)$ with respect to the distribution $p(\params)$ can be evaluated as:
\begin{equation}\label{eq:expectation}
    \langle \mathcal{O} \rangle = \frac{\int d\params\;p(\params)\mathcal{O}(\params)}{\int d\params\;p(\params)} = \frac{\int d\altvar\;g_N(\altvar)\mathcal{O}(f(\altvar))}{\int d\altvar\;g_N(\altvar)},
\end{equation}
effectively computing $\mathcal{O}(\params)$ for each sample $\params= f(\altvar)$ and taking the average. Performing such computations from the NF offers two main advantages when compared to MCMC approaches, especially for distributions in high dimensions or those with large correlations. First, we can take samples from the trivial Gaussian distribution and apply the NF in parallel, a sharp contrast with MCMC methods where configurations are generated sequentially. This, in turn, enables the parallelization of possibly time-consuming computations, for example when calculating expectation values as in Eq.~\eqref{eq:expectation} with expensive models. Second, there is no auto-correlations between samples in the Gaussian sampling process, in contrast with MCMC methods, where auto-correlation between neighboring samples is unavoidable for any distributions in any dimensions. As will be discussed in Sec.~\ref{sec:demonstration}, a sequence of samples generated by a MCMC algorithm can possess large auto-correlation lengths even for a several-dimensional distribution. 

Although an exact NF should exist for reasonably well behaved distributions~\cite{9089305}, finding such a map is often not a feasible task especially in high dimensions. Therefore, in practice, we only aim to construct an approximate NF
by relaxing the condition in Eq.~(\ref{eq:nf_exact}) and look for a function $f: \altvar \rightarrow \params$ that satisfies
\begin{equation}
    d\altvar \;g_N(\altvar) = d\params \; p'(\params) \approx d\params \; p(\params)\;\text,
\end{equation}
 where the level of the approximation is measured by a distance between two distributions, such as the Jeffreys divergence (see Eq.~(\ref{eq:Jdiv}) below), which is what we use in this study. 

When using such an approximate NF to sample from the original distribution $p(\params)$, a correction via reweighting could be used to improve the approximation. Simply, we generate Gaussian samples, apply the NF and correct their weight by the ratio $p(\params)/p'(\params)$. In the same way, to compute the expectation values using reweighting, we take the ratio:
\begin{equation}\label{eq:reweight}
    \langle \mathcal{O} \rangle_{p} = \frac{\Big\langle \frac{p(\params)}{ p'(\params)} \mathcal{O}(\params) \Big\rangle_{p'}}{\Big\langle \frac{p(\params)}{p'(\params)} \Big\rangle_{p'}}
\end{equation}
where the subscript $p'$ means that the expectation value is computed according to $p'(\alpha)$. For example, 
\begin{equation}
    \Big\langle \frac{p(\params)}{p'(\params)} \mathcal{O}(\params) \Big\rangle_{p'} = \frac{\int d\params\;p(\params)\mathcal{O}(\params)}{\int d\params\;p'(\params)}\;\text.
\end{equation}
The effective sampling size of the estimator Eq.~(\ref{eq:reweight}) is given by $1/\langle (p/p')^2\rangle_{p'} \in \left[0,1\right]$,
which takes the value 1 for a perfectly trained NF~\cite{Vaitl_2022}. 
In practice, as we demonstrate in Sec.~\ref{sec:demonstration}, a well-trained NF can closely reproduce the true posterior density, in which case the reweighing is not necessary when the NF is deployed. For the training of the NF we do use reweighing to accurately evaluate the loss function, as we explain next.

\subsection{Machine learning}\label{subsec:ml}

The construction of an approximate normalizing flow (NF) for the posterior density can be mapped onto a machine learning problem straightforwardly (see, e.g.,~\cite{Albergo:2019eim,Nicoli:2020njz,Lawrence:2021izu} for applications in lattice field theory). A neural network (NN) can be trained to represent the NF so that when it is applied to samples from the Gaussian distribution, the mapped samples approximately reconstruct the target density. The use of machine learning is motivated for the following reasons. First, since a well-designed NN is able to represent a large family of functions efficiently, it has potential for representing a good approximation to exact NFs for various posterior densities with a rather small number of network parameters. Second, once an approximate NF is obtained, it can be stored cheaply --- it costs much smaller memory to store the trained NN than keeping millions of samples of the model parameters. Furthermore, a NF enables us to access more number of samples from the posterior density. In the rest of the section, we detail our framework for building and deploying the NF through a machine learning framework.

We begin by preparing the training data for which a NF is trained. We take a supervised learning approach and prepare the data by sampling $N_t$ numbers of $\params$ via MCMC from the posterior density $p(\params)$ and storing the pairs $\{\params_i, p(\params_i)\}$ ($i=1,\cdots, N_t$). The quantity and quality of the training data directly affects the quality of the resulting NF. In particular, it is crucial that the training data covers the range of $\params$ where $p(\params)$ has appreciable support. 

The training stage is where various choices can be made for the neural network architecture, the loss function, and the training procedure such as the optimizer, the learning rate and its schedule, and the batch size. For the neural network, we use a 6-layer Real NVP (non-volume preserving) \cite{DBLP:journals/corr/DinhSB16} followed by a layer which scales and shifts each output of the Real-NVP. Schematically, the entire NN is:
\begin{equation}\label{eq:nn}
    f(\altvar) = L\circ \left( A_o \circ A_e \right)^6 (\altvar)
\end{equation}
where $A_e$ and $A_o$ denote the affine coupling layers with even and odd masking respectively. The final layer $L$, parameterized by two $N$-dimensional vectors $\boldsymbol{a}$ and $\boldsymbol{b}$, acts on each component of the output of Real NVP separately as $L(\omega_j;\boldsymbol{a},\boldsymbol{b}) = a_j \omega_j + b_j$. A great advantage of this NN is that the inverse map $f^{-1}$, and the determinant of the Jacobian of $f$ and $f^{-1}$ can be computed very efficiently. Prior to the training, the NN is initialized so that it represents an approximate Gaussian fit to the train data. For more details of the NN's architecture and its initialization, see Appendix~\ref{app:nn}.

For the loss function, we use the symmetrized Kullback–Leibler divergence, called Jeffrey's divergence~\cite{jeffreys1946invariant}: 
\begin{equation}\label{eq:Jdiv}
    D_J(p,p') = \int d\params\;\left(\tilde p(\params)\log\frac{p(\params)}{p'(\params)} + \tilde p'(\params)\log\frac{p'(\params)}{p(\params)} \right) \;\text.
\end{equation}
In our context, $p(\params)$ and $p'(\params)$ are the exact density and the density induced by the NF, respectively, while $\tilde p(\params)$ and $\tilde p'(\params)$ are those normalized to unity. Noting that the train data is sampled according to the true probability density $p(\params)$, to compute the first term in the loss function, we simply sample uniformly a subset $N_b$ (batch) of the training data ($N_b\leq N_t$) and compute the expectation value:
\begin{equation}
    \bigg\langle \log\frac{p(\params)}{p'(\params)} \bigg\rangle_p \approx \sum_{k=1}^{N_b}\log\frac{p(\params_k)}{p'(\params_k)} \;\text.
\end{equation}
The subscript $p$ means that the expectation value is computed according to $p(\params)$, therefore the samples $\params_k$ on the right hand side are sampled from $p$.
The value of the NF-induced distribution for a given sample $\params_k$ is computed as:
\begin{equation}
    p'(\params_k) = \det\left(\frac{\partial \altvar}{\partial \params}\right)g_N(\altvar_k),
\end{equation}
by using the inverse of the NF, $f^{-1}$, and obtaining $\altvar_k=f^{-1}(\params_k)$. Given that we take the supervised learning approach, the computation of the second term in Eq.~\eqref{eq:Jdiv}, which is the expectation value with respect to the NF-induced distribution, requires reweighting:
\begin{equation}
    \bigg\langle \log\frac{p'(\params)}{p(\params)} \bigg\rangle_{p'} = \frac{\big\langle \frac{p'(\params)}{p(\params)}\log\frac{p'(\params)}{p(\params)} \big\rangle_p}{\big\langle \frac{p'(\params)}{p(\params)}\big\rangle_p }\;\text.
\end{equation}
By combining the two terms, at each training step, after $N_b$
samples
are uniformly taken from the train data, the loss function is approximated by:
\begin{equation}\label{eq: approxJdiv}
    D_J(p,p')\approx
    \sum^{N_b}_k\log\frac{p(\params_k)}{p'(\params_k)} + \frac{\sum_k^{N_b}\frac{p'(\params_k)}{p(\params_k)} \log\frac{p'(\params_k)}{p(\params_k)}}{\sum_k^{N_b} \frac{p'(\params_k)}{p(\params_k)}}\text.
\end{equation}
Unlike the commonly-used Kullback–Leibler divergence, Jeffrey's divergence takes its minimal value zero only when the exact NF is represented by the neural networks, i.e. $p=p'$, regardless of the normalization of the posteriors. This feature of Jeffrey's divergence is suitable for the Bayesian framework since we typically do not know the normalization of the posterior, i.e, the evidence. 

For the training hyperparameters, in this study, we use \textsc{Adam}~\cite{kingma2017adam} optimizer, and fix the learning rate to $10^{-3}$ throughout the training process. At each training step, $N_b=10^3$ samples are randomly taken from the train data to estimate Jeffrey's loss function, and the NN parameters are optimized accordingly. We do not use any further techniques for training deep neural networks. With our algorithm, we observed a consistent improvement in the loss value as we increased the number of layers in Real NVP up to 6 for the study of the relativistic mean field model in Sec.~\ref{sec:rmf}. 

After we have trained the NN we can directly obtain uncorrelated parameter samples from the approximated posterior density. For computing expectation values as in Eq.~\eqref{eq:expectation}, we could furthermore correct for the slight deviation from the true posterior distribution, as discussed in the previous section, by computing the ratio $p(\params_k)/p'(\params_k)$ at each sample $k$. 

\section{Applications}\label{sec:demonstration}
\begin{figure*}[!ht]
    \includegraphics[width=1\linewidth]{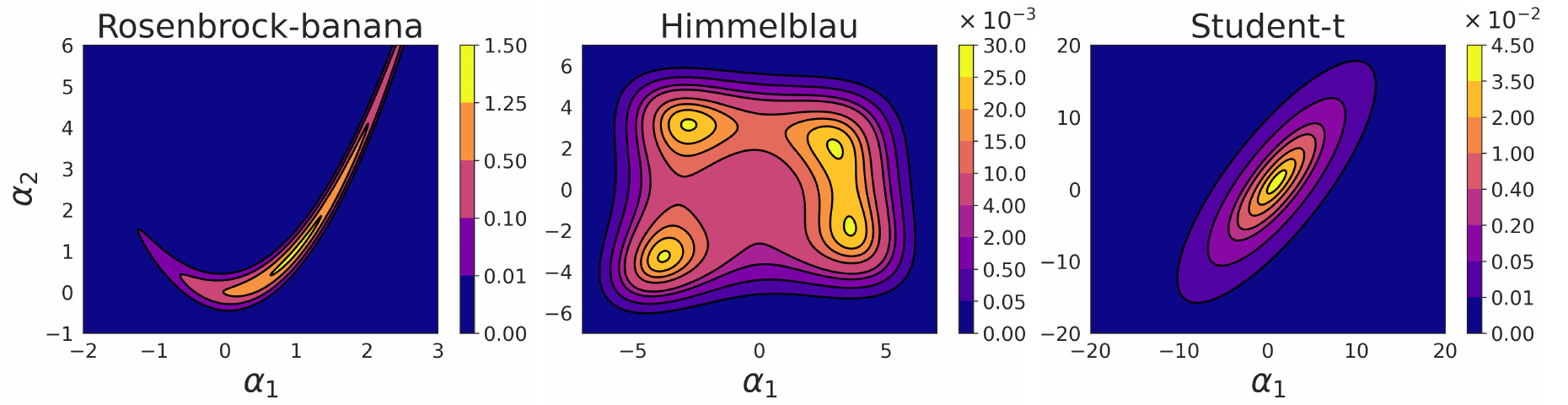 } 

    \caption{Two-dimensional distributions $p(\alpha_1,\alpha_2)$ we consider in this work for testing the NF framework: Rosenbrock-banana \cite{rosenbrock1960automatic} (left), Himmelblau \cite{Himelblau72} (center), and  Student-t \cite{gelman1995bayesian} (right). The colors in each legend indicate the numerical value of each normalized density along the respective contours.}
    \label{fig:dist_test}
\end{figure*}

To demonstrate the algorithm described in the previous section, we analyze three two-dimensional distributions as tests: the Rosenbrock's banana function \cite{rosenbrock1960automatic}, an exponentiated Himmelblau distribution \cite{Himelblau72}, and the Student's t-distribution \cite{gelman1995bayesian}. Afterwards, we focus on a realistic case by learning the NF for the distribution obtained in~\cite{giuliani2023bayes}, which was constructed by performing a Bayesian calibration of a relativistic mean field nuclear model. We used the JAX library Equinox~\cite{kidger2021equinox} for the implementation of the neural network in this work. All the data, the codes, and the trained NF used to produce these results are publicly available \href{https://gitlab.com/yyamauchi/rbm_nf}{here}.

\subsection{2-dimensional test distributions}
The three distributions we selected for testing the NF highlight three challenging features: the Rosenbrock's banana function has a highly correlated non-linear structure, the exponentiated Himmelblau distribution is multi-modal, and the Student's t-distribution has heavy tails. These test distributions are shown in Fig.~\ref{fig:dist_test}.

For each distribution, we generated a single MCMC chain for the training data. In the chain, 1000 steps were taken first for burn-out and the acceptance rate was adjusted to be in the range $\left[0.3, 0.5\right]$. Then one sample per 100 MCMC steps was taken to partially reduce the auto-correlations. A total of $N_t=10^5$ samples were collected in each case and used in the training of the respective NFs.

The Rosenbrock-banana distribution, up to a normalization constant, can be written as:
\begin{equation}\label{eq:banana}
    p_b(\paramsOneD_1, \paramsOneD_2) = e^{-(\paramsOneD_1-c_1)^2-c_2(\paramsOneD_1^2-\paramsOneD_2)^2}\;\text.
\end{equation}
We choose to work with the coefficients $c_1=1.0$ and $c_2 = 20.0$. The auto-correlation length in the MCMC chain was estimated to be $l\sim [500-1000]$ from the correlation function in Eq.~(\ref{eq:ac}) --- it is large due to the high correlation structure between the two parameters. This structure is apparent in Fig.~\ref{fig:dist_test}: the distribution is not localized within a convex boundary but rather extends and curves across the parameter space.

The exponentiated Himmelblau distribution, up to a normalization constant, can be written as:
\begin{equation}\label{eq:himmelblau}
    p_h(\paramsOneD_1, \paramsOneD_2) = e^{-\frac{1}{c_3}[(\paramsOneD_1^2+\paramsOneD_2-c_1)^2 + (\paramsOneD_1+\paramsOneD_2^2-c_2)^2]}.
\end{equation}
We choose to work with the coefficients $c_1=11.0$, $c_2=7.0$, and $c_3=100.0$. The distribution has, as is shown in Fig.~\ref{fig:dist_test}, four peaks that are connected by a non-vanishing density. Therefore a MCMC chain could reasonably explore those four regions and offer sufficient training data. The auto-correlation length in the MCMC sampling was estimated to be $l\sim 100$ from the correlation function in Eq.~(\ref{eq:ac}).

The bivariate Student-t distribution, up to a normalization constant, can be written as:
\begin{equation}
    p_s(\paramsOneD_1, \paramsOneD_2) = \left[1 + \frac{y^{T} M y}{c}  \right]^{-(2+c)/2}.
\end{equation}
We choose to work with the mean and scaled variance matrix:
\begin{equation}\label{eq:student-t}
y = \begin{pmatrix}
\paramsOneD_1 - 1.0 \\
\paramsOneD_2 - 1.0 
\end{pmatrix}
 \;\;\mathrm{and}\;\;
    M = \begin{pmatrix}
4.0 & 4.8 \\
4.8 & 9.0 
\end{pmatrix}^{-1}.
\end{equation}
The parameter $c$ determines how quickly the density decays to zero, with smaller values creating heavier tails. We choose $c=3.$ in our study to have very heavy tails, creating a distribution with only two well defined moments (means and variances). As can be seen in Fig.~\ref{fig:dist_test}, the density extends over a big region of the 2D parameter space, with an appreciable linear correlation between them. The auto-correlation length in the MCMC sampling was estimated to be $l \sim 100$ from the correlation function in Eq.~(\ref{eq:ac}).

\begin{figure*}[!htb]

    \includegraphics[width=1\linewidth]{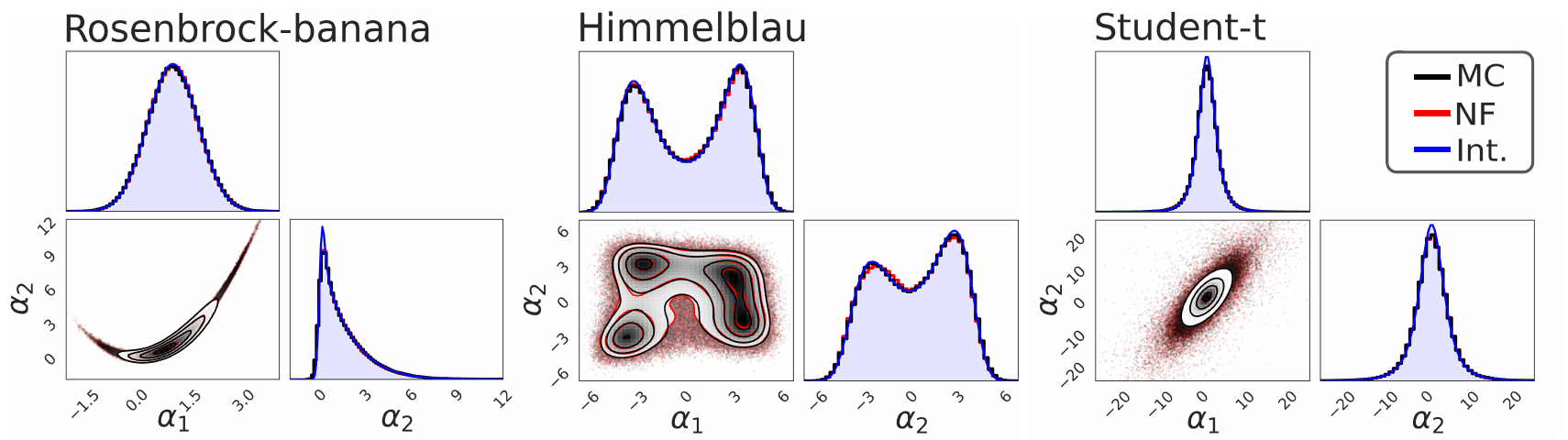 } 

    \caption{Comparison between the samples obtained by MCMC (black) and NF (red) for the three test distributions $p(\alpha_1,\alpha_2)$: Rosenbrock-banana (left),  Himmelblau (center), and Student-t (right). For comparison, the original distributions defined in Eqs.~\eqref{eq:banana}, \eqref{eq:himmelblau}, and \eqref{eq:student-t} are also plotted (blue) and labeled as ``Int." to stand for numerical integration. The NF samples reproduce all the features of the corner plot of the MCMC, overlapping well enough to make it hard to distinguish between them.}
    \label{fig:corner_test}
\end{figure*}

The $N_t$ MCMC training samples together with the same number of samples obtained from the trained NFs are compared in Fig.~\ref{fig:corner_test}. In all cases, the NFs successfully reproduced key features of the test cases: the non-linear high correlation, the multi-modal structure and the heavy tails.

To quantitatively assess the performance of the NF Table~\ref{table:test} presents the calculations of the first two moments (means and covariances) of the three distributions, as well as the final value of the loss function \eqref{eq: approxJdiv} during training. In all cases, the NFs obtained very small values for Jeffrey's divergence, implying that the effective sampling size be close to one. Therefore all results, means and covariances, from the NFs were computed with respect to the approximate density $p'$, i.e., without reweighting. In all cases the obtained means and covariances from the NF are in good agreement with the estimate from the train data (MCMC in the table). Both calculations are also in agreement with those obtained by numerical integration (Int.), with the Student-t variances being moderately under-estimated by the sampling methods.

\begin{table}[]
\begin{tabular}{cc|c|c|c|}
\cline{3-5}
                                                                 &                       & Banana                & Himmelblau            & Student-t             \\ \hline
\multicolumn{1}{|c|}{\multirow{3}{*}{$\langle \alpha_1\rangle$}} & Int.                  & 1.00                 & 0.110                 & 1.00                 \\ \cline{2-5} 
\multicolumn{1}{|c|}{}                                           & MC                    & 1.004(8)              & 0.124(8)              & 1.00(1)             \\ \cline{2-5} 
\multicolumn{1}{|c|}{}                                           & NF                    & 1.002(2)              & 0.11(1)             & 0.99(1)             \\ \hline
\multicolumn{1}{|c|}{\multirow{3}{*}{$\langle \alpha_2\rangle$}} & Int.                  & 1.50                 & 0.230                 & 1.00                 \\ \cline{2-5} 
\multicolumn{1}{|c|}{}                                           & MC                    & 1.51(2)              & 0.238(7)              & 1.00(3)             \\ \cline{2-5} 
\multicolumn{1}{|c|}{}                                           & NF                    & 1.503(5)              & 0.211(8)              & 1.00(2)             \\ \hline
\multicolumn{1}{|c|}{\multirow{3}{*}{$\text{var} (\alpha_1)$}}   & Int.                  & 0.500                 & 8.96                 & 11.9                 \\ \cline{2-5} 
\multicolumn{1}{|c|}{}                                           & MC                    & 0.505(9)              & 8.92(2)             & 11.3(4)           \\ \cline{2-5} 
\multicolumn{1}{|c|}{}                                           & NF                    & 0.500(2)              & 8.97(2)             & 11.2(2)           \\ \hline
\multicolumn{1}{|c|}{\multirow{3}{*}{$\text{var} (\alpha_2)$}}   & Int.                  & 2.53                 & 6.39                 & 26.8                 \\ \cline{2-5} 
\multicolumn{1}{|c|}{}                                           & MC                    & 2.5(1)                & 6.42(2)             & 26(1)           \\ \cline{2-5} 
\multicolumn{1}{|c|}{}                                           & NF                    & 2.54(2)               & 6.39(2)             & 25.8(7)           \\ \hline
\multicolumn{1}{|c|}{\multirow{3}{*}{$\text{cov} (\alpha_1\alpha_2)$}}  & Int.                  & 1.00               & 0.225                 & 14.4                   \\ \cline{2-5} 
\multicolumn{1}{|c|}{}                                           & MC                    & 1.01(3)              & 0.20(3)             & 13.7(6)           \\ \cline{2-5} 
\multicolumn{1}{|c|}{}                                           & NF                    & 1.005(6)              & 0.22(3)             & 13.4(3)           \\ \hline
\multicolumn{1}{|l}{Jeff. Div.} & \multicolumn{1}{l|}{} & \multicolumn{1}{l|}{$1.0\times10^{-5}$} & \multicolumn{1}{l|}{{$3.8\times 10^{-3}$}} & \multicolumn{1}{l|}{$6.0\times 10^{-5}$} \\ \hline

\end{tabular}

\caption{Comparison between the first two moments for each of the three test distributions we considered. These moments are calculated using numerical integration in python (Int.), as well as samples from Markov-Chain Monte Carlo (MC), and Normalizing Flow (NF) in each case. The last digit in parentheses for MC and NF shows the associated standard deviation on the calculation obtained by data blocking and statistical bootstrap (see, e.g.~\cite{Gattringer:2010zz} for the methods). Jeffreys divergence between MCMC and NF -the loss function used to train the NF in Eq.~\eqref{eq: approxJdiv}- is also shown in the last row.}\label{table:test}

\end{table}

\subsection{RMF}\label{sec:rmf}

Having tested the NF framework on the three known 2-dimensional distributions, we proceed to study a realistic Bayesian posterior without an explicit functional form \cite{giuliani2023bayes}. This posterior constrains the 8 parameters $\params$ from a relativistic mean field model \cite{Todd:2003xs} used to calculate properties of finite nuclei and neutron stars. The original data consisted of $5\times10^6$ MCMC samples that were obtained from 8 walkers, each with a 100,000 burn-out period. Obtaining such a large number of samples in the original work was possible thanks to the use of a reduced order model, since each full evaluation of the original model could take several minutes. In constructing the train data for the NF, we took one sample per 100 MCMC steps in each chain, resulting in a total of $N_t=5\times10^4$ samples.

\begin{figure*}[!htbp]
    \centering{\includegraphics[width=0.95\textwidth]{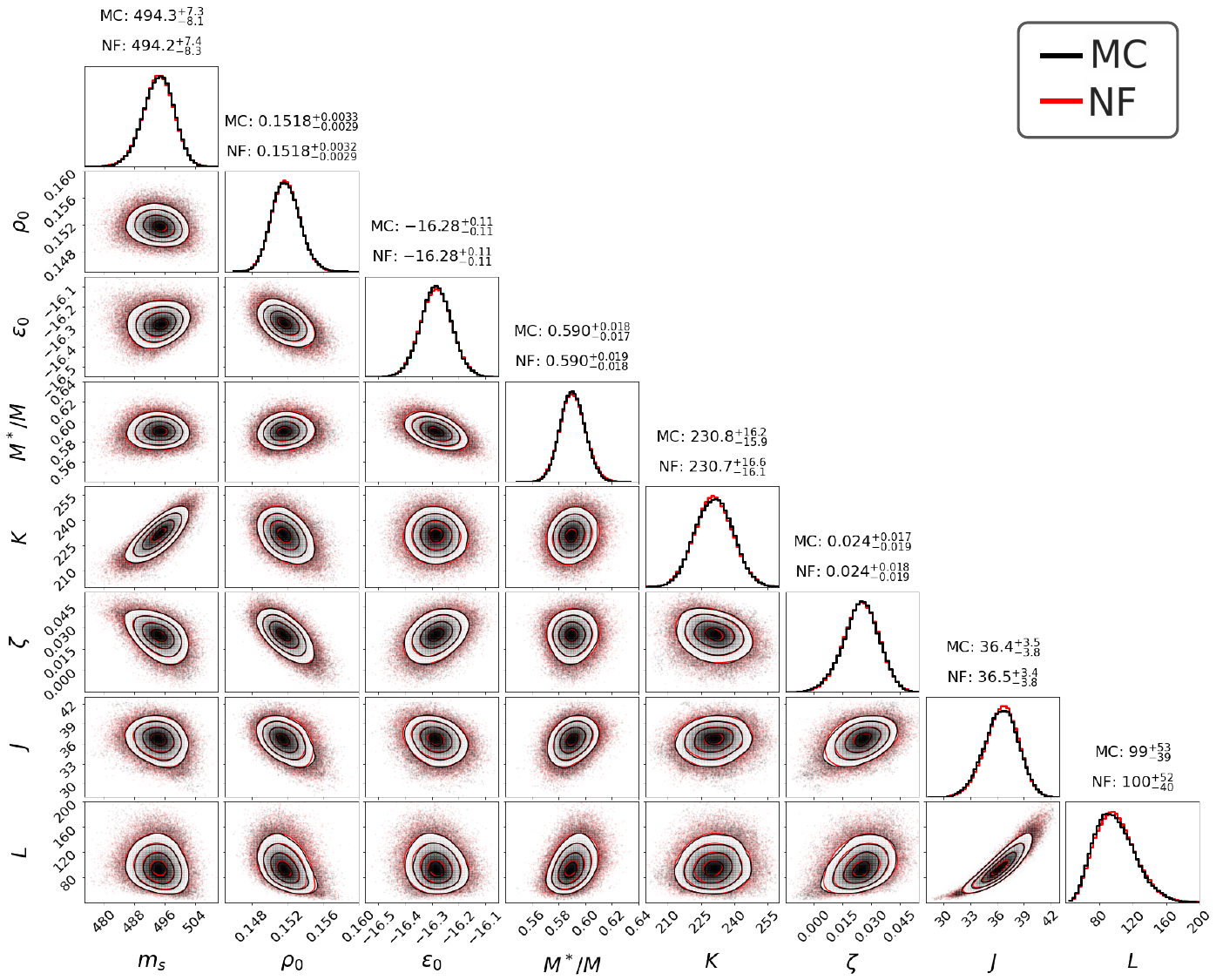}}
    \caption{Comparison between the samples obtained by MCMC (black) and NF (red) for the relativistic mean field Bayesian calibration \cite{giuliani2023bayes}. Similar to the situation in Fig.~\ref{fig:corner_test}, the NF samples reproduce all the features of the corner plot of the MCMC, overlapping well enough to make it hard to distinguish between them. The distinction is most noticeable in the $J$ and $L$ univariate plots. }
      
    \label{fig:MCMC}
\end{figure*}

The NF for this model was built following the steps described in Sec.~\ref{subsec:ml}. We initialized the neural network in Eq.~\eqref{eq:nn} to an approximate Gaussian fit with means and covariances estimated from the train data. The NN was then trained to minimize the loss function in Eq.~\eqref{eq: approxJdiv} by estimating it using $N_b=10^3$ randomly-drawn samples from the train data at each training step. As for the optimizer, we used again \textsc{Adam} with the learning rate of $10^{-3}$. 

Jeffrey's divergence was $\sim 10$ at the initialization (with an approximate Gaussian fit) and reduced to $2\times10^{-3}$ after training. Fig.~\ref{fig:MCMC} shows the comparison between the $5\times10^4$ training MCMC samples, and the same number of samples obtained from the trained NF. Qualitatively, the NF reproduces all the features from the distribution estimated from the MCMC samples, in particular the non-linear correlation between the symmetry energy $J$ and its slope $L$. For a quantitative assessment, we show on top of the diagonal univariate plots of Fig~.\ref{fig:MCMC} for both the MCMC and the NF samples the calculated means and the $95\%$ confidence interval around them.  In all cases the NF calculation matches the MCMC one almost perfectly.

\section{Conclusions and Outlook}\label{sec:conclusion}

We have presented an efficient method to accurately reconstruct Bayesian posterior distributions, through the use of NFs approximated through neural networks, for the purpose of further sampling and deployment. The NF map is learned by a conventional machine learning procedure by minimizing a loss function -Jeffreys divergence in Eq.~\eqref{eq:Jdiv}- over training data in the form of a finite number of MCMC samples from the posterior. The NF framework was able to successfully learn the three test distributions with known forms we explored, even in the presence of high non-linear correlations,  multi-modality, and heavy tails. In the more realistic scenario involving the Bayesian posterior of the parameters of a physical model~\cite{giuliani2023bayes}, the NF reproduced the target distribution with a similar level of accuracy.

Once trained, the deployed NF enables us to obtain uncorrelated samples from the posterior by directly sampling the Gaussian distribution. This in turns can speed-up the  evaluation of Bayesian integrals -such as expectation values- by parallelizing the calculation over several independent nodes. Furthermore, the trained NF requires only a small amount of memory to be stored, facilitating both the dissemination of results from calibration efforts, as well as increasing the reproducibility of studies based on those results. The NF could also provide an alternative to MCMC based methods for the construction of the posterior from the beginning, bypassing the need to compute long correlated chains. Such self-learning approaches will also enable us to overcome issues with supervised learning and improve the quality of NFs.

We believe that the implementation of NFs could become a standard procedure in the workflow of distributing and updating results derived from Bayesian calibrations. By using trained NFs from previous calibrations as the starting point for updating posteriors distributions when new data becomes available, this framework could enable a continuous calibration of complex models. In this way, the full Bayesian philosophy is embraced by using all previous information as prior knowledge, and continuously updating it as new vetted data is made available.
By deploying the trained NFs in turn, researchers can directly draw samples from the distributions for their own purposes.
Furthermore, each trained NF can be tagged and distributed alongside metadata defining the training process to provide a traceable and reproducible workflow.

\begin{acknowledgements}

Y.Y.~is grateful to Scott Lawrence for many useful discussions and suggestions for the machine learning tools and techniques used in this work. P.G. and K.G.~are grateful to Edgard Bonilla and Frederi Viens for useful discussions on the selection of distributions and on the interpretation of the results. Y.Y.~is supported by the INT's U.S. Department of Energy grant No.~DE-FG02-00ER41132. This work has also been supported by the National Science Foundation
CSSI program under award number 2004601 (BAND collaboration) and the U.S. Department of Energy under award number DE-SC0023175 (NUCLEI SciDAC-5 collaboration).
This work was supported in part through computational resources and services provided by the Institute for Cyber-Enabled Research at Michigan State University.
\end{acknowledgements}

\bibliographystyle{apsrev4-1}
\bibliography{nf}

\appendix

\section{Neural Network}\label{app:nn}
The structure of the neural network used in this study is, as in Eq.~(\ref{eq:nn}),
\begin{equation}
    f(\altvar) = L \circ \left( A_o \circ A_e \right)^6 (\altvar) \;\text.
\end{equation}
The first block that the input $\altvar$ goes through is a 6-layer Real-NVP (non-volume preserving) \cite{DBLP:journals/corr/DinhSB16} neural network. Each layer consists of two affine coupling layers --- $A_e$ with the even masking $b_e$ followed by $A_o$ with the odd masking $b_o$. The affine coupling layer is constructed as
\begin{equation}\label{eq:affine}
    A(\altvar; b) = b\odot\altvar + (1-b)\odot\left[\altvar\odot \exp (s(b \odot \altvar))+ t( b\odot \altvar)\right]
\end{equation}
where $\odot$ is the element-wise product. The masking patterns are
\begin{eqnarray}
    b_e = \left[ 0,1,0,1,\cdots\right] \;\;\;\mathrm{and}\;\;\; b_o = \left[ 1,0,1,0,\cdots\right]
\end{eqnarray}
We used a single linear layer for the network $t$ and an MLP with one hidden layer and the hyperbolic tangent activation function for the network $s$. Note that when the networks $s$ and $t$ output zero, the affine coupling layer acts as the identity function. This can be arranged by setting all parameters in $t$ and the final linear layer in $s$ to zero. 

The second block of the neural network is a simple linear transformation that acts on each output of the Real-NVP separately:
\begin{equation}
    L(\altvar;\boldsymbol{a},\boldsymbol{b}) = \boldsymbol{a} \odot \altvar + \boldsymbol{b}\;\text.
\end{equation}
The purpose of this layer is to encode a rough estimate of the mean and variance of the parameters $\params$, so the Real-NVP deals only with numbers of order unity. This implementation not only makes the training efficient but also was necessary for stabilizing the evaluation of Jeffreys' divergence. 

Prior to the training of the neural network, is it initialized so that it yields an approximate Gaussian fit to the train data. This is arranged by initializing the Real-NVP as the identity, and setting the coefficients in $L$ by the mean and variance approximated from the train data. The parameters in the first layer of $s$ were set randomly from  the Gaussian distribution.

In evaluating the loss function Eq.~(\ref{eq:Jdiv}), it is important that the inverse of the NF and the determinant of the Jacobian is computed efficiently for this neural network via convolution. The inverse function of an affine coupling layer is
\begin{equation}
A^{-1}(\params; b) = b\odot \params + (1-b) \odot (\params-t(b\odot \params)) \odot \exp(-s(b\odot \params))
\end{equation}
Therefore the inverse of the normalizing flow is
\begin{equation}
    f^{-1}(\params) =  \left( A^{-1}_e \circ A^{-1}_e \right)^6 \circ L^{-1}(\boldsymbol{a},\boldsymbol{b}) (\params) \;\text.
\end{equation}
The determinant of the Jacobian of the map $f$ Eq.~(\ref{eq:nn}) can be computed very efficiently by combining that of the single affine coupling layer 
\begin{eqnarray}
    \log\det\left(\frac{\partial A(\altvar;b)}{\partial\altvar}\right) = (1-b) \cdot s(b\odot \altvar) 
\end{eqnarray}
and likewise for the inverse map (see~\cite{DBLP:journals/corr/DinhSB16} for further details).

\end{document}